The validation of (advanced) bibliometric indicators through peer assessments:

A comparative study using data from InCites and F1000

Lutz Bornmann* & Loet Leydesdorff[$]

* Division for Science and Innovation Studies, Administrative Headquarters of the Max Planck Society, Hofgartenstr. 8, 80539 Munich, Germany. E-mail: bornmann@gv.mpg.de

$ Amsterdam School of Communication Research (ASCoR), University of Amsterdam, Kloveniersburgwal 48, 1012 CX Amsterdam, The Netherlands. E-mail: loet@leydesdorff.net

**Abstract**

The data of F1000 provide us with the unique opportunity to investigate the relationship between peers' ratings and bibliometric metrics on a broad and comprehensive data set with high-quality ratings. F1000 is a post-publication peer review system of the biomedical literature. The comparison of metrics with peer evaluation has been widely acknowledged as a way of validating metrics. Based on the seven indicators offered by InCites, we analyzed the validity of raw citation counts (Times Cited, 2nd Generation Citations, and 2nd Generation Citations per Citing Document), normalized indicators (Journal Actual/Expected Citations, Category Actual/Expected Citations, and Percentile in Subject Area), and a journal based indicator (Journal Impact Factor). The data set consists of 125 papers published in 2008 and belonging to the subject category cell biology or immunology. As the results show, Percentile in Subject Area achieves the highest correlation with F1000 ratings; we can assert that for further three other indicators (Times Cited, 2nd Generation Citations, and Category Actual/Expected Citations) the "true" correlation with the ratings reaches at least a medium effect size.

**Key words**

Advanced bibliometric indicators; Peer review; F1000; InCites

# 1  Introduction

Research quality is a complex attribute that takes into account various factors such as importance, originality, rigour, elegance, and scientific impact (Council of Canadian Academies, 2012). Since there is no mathematical formula that can quantify the "quality" of a paper (Figueredo, 2006), a set of different citation-based indicators are used in bibliometrics to measure quality. Overviews of current indicators are provided by Rehn, Kronman, and Wadskog (2007) and Vinkler (2010). The simplest indicator is the number of times a paper has been cited (the times cited indicator). Since scientific fields (and also subfields) are characterized by different expected citation rates, citation counts across different fields (and also subfields) cannot be directly compared (Bornmann & Daniel, 2009; Bornmann, Mutz, Marx, Schier, & Daniel, 2011).

In order to overcome this problem in citation analysis, bibliometricians have proposed several methods of field-normalization. Field-normalized citation rates of papers published in different fields are directly comparable in terms of the papers' impact. The most well-known field-normalizing techniques for a single paper are: (1) Each paper's citation count is divided by the mean citation rate of papers published in the same field (and in the same year) as the paper in question (the field-based reference set); (2) Each paper's citation counts is divided by the mean citation rate of papers published in the same journal (and in the same year) as the paper in question (the journal-based reference set); (3) Percentiles offer an alternative to the mean-based approaches in (1) and (2). A percentile is a value below which a certain proportion of observations fall. Using a reference set it is possible to determine, for example, whether a paper in question belongs to the 1% or 10% of the most-cited papers in the (field-based) reference sets or not.

Hitherto, it is not clear which of the different techniques should be preferred in citation analysis. Although many studies have compared the various techniques in the past they could not come to a satisfying answer. In these studies, one argumentatively justified why one indicator should be preferred against another one (e.g. Bornmann, 2010) or statistical methods (e.g. factor analysis) were used to combine many correlated metrics into a smaller number of underlying "quality" dimensions (e.g. Bollen, Van de Sompel, Hagberg, & Chute, 2009; Bornmann, Mutz, & Daniel, 2008; Leydesdorff, 2009). However, these studies did not compare the different citation-based metrics with an independent and expert-based indicator of quality.

The comparison of metrics with peer evaluation has been widely acknowledged as a way of validating metrics (Garfield, 1979; Kreiman & Maunsell, 2011). Several publications investigating the relationship between citations and Research Assessment Exercise (RAE) outcomes report considerable relationships in several subjects like biological science, psychology, and clinical sciences (Butler & McAllister, 2011; Mahdi, d'Este, & Neely, 2008; McKay, 2012; Smith & Eysenck, 2002). Similar results were found for the Italian research assessment exercise, called Valutazione Triennale della Ricerca (VTR): "The correlation strength between peer assessment and bibliometric indicators is statistically significant, although not perfect. Moreover, the strength of the association varies across disciplines, and it also depends on the discipline internal coverage of the used bibliometric database" (Franceschet & Costantini, 2011, p. 284). An overview of papers that report a close relationship between peer ratings or editorial decisions, respectively, at single journals and bibliometric metrics investigated can be found in Bornmann (2011).

Since most of these studies used only simple metrics, which were not field-normalized, or restricted to single journals, a study considering advanced bibliometric indicators on a larger scale is needed. In this study, we calculate Spearman's rank order correlations between different (field-normalized) citation impact values and the peer ratings gathered in Faculty of 1000 (F1000, http://f1000.com/) for published papers. For F1000, researchers review and rate biomedical papers in their specialist areas (Li & Thelwall, 2012).

In our opinion, the bibliometric metric with the highest correlation coefficient should be preferred to the other metrics in research evaluation. For example, as Wouters and Costas (2012) formulate: "the data and indicators provided by F1000 are without doubt rich and valuable, and the tool has a strong potential for research evaluation, being in fact a good complement to alternative metrics for research assessments at different levels (papers, individuals, journals, etc.)" (p. 14).

## 2 Methods

### 2.1 Peer ratings provided by F1000

F1000 is a post-publication peer review system of the biomedical literature (papers from medical and biological journals). Papers are selected by a peer-nominated global "Faculty" of leading scientists and clinicians who then rate them and explain their importance (F1000, 2012). This means that only a restricted set of papers from the medical and biological journals covered is reviewed, and most of the papers are actually not (Kreiman & Maunsell, 2011; Wouters & Costas, 2012).

The Faculty nowadays numbers more than 5,000 experts worldwide, assisted by 5,000 associates, which are organized into more than 40 subjects. On average, 1500 new recommendations are contributed by the Faculty each month (F1000, 2012). Faculty members can choose and evaluate any paper that interests them; however, "the great majority pick papers published within the past month, including advance online papers, meaning that users can be made aware of important papers rapidly" (Wets, Weedon, & Velterop, 2003, p. 254). Although many papers published in popular and high-profile journals (e.g. *Nature*, *New England Journal of Medicine*, *Science*) are evaluated, 85% of the papers selected come from specialized or less well-known journals (Wouters & Costas, 2012). "Less than 18 months since Faculty of 1000 was launched, the reaction from scientists has been such that two-thirds of top institutions worldwide already subscribe, and it was the recipient of the Association of Learned and Professional Society Publishers (ALPSP) award for Publishing Innovation in 2002 (http://www.alpsp.org/about.htm)" (Wets, et al., 2003, p. 249).

The papers are rated by the members as "Recommended," "Must read" or "Exceptional" which is equivalent to scores of 6, 8, or 10, respectively. The FFa is calculated from the highest rating awarded by a member (6, 8, or 10) plus an increment for each additional rating from other members (1, 2 or 3 for "Recommended," "Must Read" or "Exceptional," respectively). For example, a single paper that has been evaluated by three Faculty members, who scored it "Exceptional", "Must Read" and "Recommended," will have a FFa of 13. The calculation is: 10 ("Exceptional") + 2 ("Must Read") + 1 ("Recommended") = 13 (Li & Thelwall, 2012). The FFa is listed with the bibliographic information of a paper at the F1000 web site.

**2.2    Data set for the study**

Two sets with papers published in 2008 were downloaded for this study from InCites (Thomson Reuters): 2,657 papers belonging to the subject category cell biology (InCites$^{TM}$ Thomson Reuters, 2012b) and 2,547 papers belonging to immunology (InCites$^{TM}$ Thomson Reuters, 2012a). InCites (http://incites.thomsonreuters.com/) is a web-based research evaluation tool allowing assessment of the productivity and citation impact of researchers, institutions, and countries. The metrics (such as the percentile for each individual publication) are generated from Web of Science (WoS, Thomson Reuters) publications from 1981 to 2010. Since we have access only to the address search dataset Germany in InCites, the downloaded papers are restricted to only those with at least one German address. All these papers were searched in f1000.com for gathering their FFa. For 125 of the total 5,204 papers (2.4%) a FFa could be retrieved.

From InCites, the following bibliometric metrics at the paper level were downloaded and correlated with FFa (the definitions of the metrics below were taken from the glossary of InCites Help):

(1) Times Cited: "Total number of citations from Web of Science (as of last InCites update)."

(2) 2nd Generation Citations: "Total number of citations received by the citing papers of a source article."

(3) 2nd Generation Citations per Citing Document: "Total number of citations received by all citing papers divided by the number of citing papers."

(4) Journal Actual/Expected Citations: The category expected citations is the "average number of citations to articles of the same document type from the same journal in the same database year. You can compare an article's citation count to this norm by forming a ratio of actual citations to expected citations – the Journal Actual/Expected Citations. A ratio greater than 1 indicates that the article's citation count is better than average"

(5) Category Actual/Expected Citations: The category expected citations is the "average number of citations received by articles of the same document type from journals in the same database year and same category (subject area). If a journal is assigned to more than one category, the category expected cites is the average for the categories. You can compare an article's citation count to this norm by forming a ratio of actual citations to expected citations – the Category Actual/Expected Citations. A ratio greater than 1 indicates that the article's citation count is better than average"

(6) Percentile in Subject Area: "The percentile in which the paper ranks in its category and database year, based on total citations received by the paper. The higher the number [of] citations, the smaller the percentile number [is]. The maximum percentile value is 100, indicating 0 citations received. Only article types article, note, and review are used to determine the percentile distribution, and only those same article types receive a percentile value. If a journal is classified into more than one subject area, the percentile is based on the subject area in which the paper performs best, i.e. the lowest value." Since this indicator is the only indicator where small values mean high citation impact, the percentile values were reversed by us using for each paper 100 minus the percentile value provided by InCites.

(7) Journal Impact Factor: "Average number of times articles from a journal published in the past two years have been cited in the JCR year. For example, a 2009 Journal Impact Factor of 4.25 means that, on average, an article published in the journal in 2007 or 2008 received 4.25 citations in 2009. The journal impact factor displayed [in InCites] is the most current journal impact factor available."

Table 1 shows the summary statistics of the bibliometric metrics and FFa for the 125 papers included in the study. As the number of papers in the table for the metrics and FFa reveal, for three metrics some missing values appear.

## 2.3    Statistical procedures

Since the distributions of the variables in Table 1 are not (approximately) normal (examined with a test for normality based on skewness and another based on kurtosis and then a combination of the two tests into an overall test statistic), the Spearman's rank-order correlation coefficient ($r_s$) is used to determine the degree to which a relationship exists between FFa and a bibliometric metric (Sheskin, 2007, test 29). The size of the correlational effects in this study is interpreted using the recommendations of Cohen (1988, Chapter 3). Many sources recommend that the $r_s$ value should be adjusted, when one or more ties in the data are present. Although this is the case for some variables in our data set (especially for FFa), we decided to calculate the unadjusted coefficients, since (1) the correction for ties is not available in the software package which we used for the statistical analysis (StataCorp., 2011) and (2) Sheskin (2007) demonstrates that the differences between the adjusted and unadjusted coefficients are very small.

Upon computing the coefficient, "it is common practice to determine whether the obtained absolute value of the correlation coefficient is large enough to allow a researcher to conclude that the underlying population correlation coefficient between the two variables is some value other than zero" (Sheskin, 2007, p. 1355). Statistical significance implies "that the outcome of a study is highly unlikely to have occurred as a result of chance" (Sheskin, 2007, p. 67). However, random sampling is a prerequisite for the process of drawing conclusions from an observed dataset to the population. The papers in the data set of this study are not a random sample of a larger population and are not selected by any known sampling method. Thus, the question which should be answered by the statistical test in this study is whether a relationship between two variables could have happened (with a decent likelihood) because of a random data-generating process or whether it is systematically linked to some key variables of interest (here: the bibliometric metrics and FFa) (Kohler & Kreuter, 2012).

We use the Bonferroni adjustment to counteract the problem of multiple comparisons which can result from the multiple correlation of FFa with several bibliometric metrics. Furthermore, confidence intervals for each correlation coefficient are calculated: Thus, we can be 95% confident (or the probability is .95) that the interval contains the "true" correlation coefficient in the underlying population (Sheskin, 2007).

## 3  Results

Figure 1 shows the Spearman's rank order correlation coefficients with 95% confidence intervals for the correlation between FFa and the seven bibliometric metrics. As the results point out there are indeed large differences between the metrics: Whereas the lowest correlation is obtained for the indicator Journal Actual/Expected Citations; the highest is revealed for Percentile in Subject Area. According to Cohen (1988), one useful way to approach an understanding of r is to compute $r^2$. The square of the coefficient is the proportion of variance in FFa which may be accounted for by the variance of a metric. Also, $r^2$ can be interpreted the other way around (as an explanation of the variance in metric), since the attribution of causality is not clear here. However, we assume that citations measure only one of three aspects of quality (defined by Martin & Irvine, 1983), namely impact. Peers can be expected to assess most likely all three aspects of quality, namely "importance" (the influence of research on the advance of scientific knowledge) and "accuracy" (how well the research has been done) in addition to "impact." Thus, it is useful to ask for the explanation of variance in FFa by the various metrics.

Whereas Journal Actual/Expected Citations explains only 1% of the variance in FFa, Percentile in Subject Area reaches a value of 20%. Although the latter metric points out the highest percentage in Figure 1, the value seems to be small against a theoretically possible value of 100%. According to Cohen (1988), however, one has to consider in the interpretation of correlation coefficients in the behavioral sciences that they are small as a rule. Coefficients above the .50-.60 range are normally encountered only when the correlations are measurements of reliability coefficients (in personality-social psychology). Against the expectation of general small coefficients in the behavioral science, Cohen (1988) interprets $r=.5$ ($r^2=25\%$) as a large, $r=.3$ ($r^2=9\%$) as a medium, and $r=.1$ as a small effect ($r^2=1\%$). Interpreted against these effect size thresholds provided by Cohen (1988), the "true" relationship between Percentile in Subject Area and FFa has at least a medium and at best a large effect size. Given that the assessment of a paper's quality by a peer consists of three aspects (impact, importance, and accuracy), one can expect that the measurement of only impact (namely citations) can explain one third of the variance in the peer ratings at the maximum ($r^2=33\%$ or $r=.57$). Using $r^2=33\%$ as the expected value, Percentile in Subject Area with an $r^2=20\%$ comes close to this value.

## 4    Discussion

Assessments by peers are the cornerstones of modern science (Ruscio, Seaman, D'Oriano, Stremlo, & Mahalchik, in press). High-quality research is the result of research and its assessment by peers. Since many years, bibliometric (citation-based) indicators have been developed as a valuable alternative to peer review (Bornmann, 2011). "Although such indices (e.g., journal impact factors) do not capture the multidimensional complexities of article quality … they are widely used proxies that do not suffer the unreliability that plagues more subjective quality assessments" (Haslam & Laham, 2010, p. 217). The bibliometric indicators should demonstrate, however, their validity as measured by the established assessments by peers: There should be a close relationship between both measures, but one should consider in this comparison that citation-based indicators measure only one aspect of research quality (its impact). Peers can additionally assess the other two aspects (accuracy and importance). Thus, one can expect that these indicators can explain only a part (or one third) of the variance in peer ratings.

Using the data of F1000 (FFa) we have the unique opportunity to investigate the relationship between peers' ratings and bibliometric metrics on a broad and comprehensive data set with high-quality ratings. Other studies using FFa data (Allen, Jones, Dolby, Lynn, & Walport, 2009; Medical Research Council, 2009; Wardle, 2010) did not have this focus of interest: correlations between FFa and (advanced) bibliometric indicators. Based on the seven indicators offered by InCites, we thus analyzed the validity of raw citation counts (Times Cited, 2nd Generation Citations, and 2nd Generation Citations per Citing Document), normalized indicators (Journal Actual/Expected Citations, Category Actual/Expected Citations, and Percentile in Subject Area), and a journal based indicator (Journal Impact Factor).

As the results show, Percentile in Subject Area achieves the highest correlation with FFa; we can assert that for further three other indicators (Times Cited, 2nd Generation Citations, and Category Actual/Expected Citations) the "true" correlation with FFa reaches at least a medium effect size. As an important reason for the relatively high correlation coefficients of the two indicators which are based on raw citation counts, we assume the selection of papers from two similar subject categories cell biology and immunology. According to InCites (InCites[TM] Thomson Reuters, 2012c) the citation indicator "Impact relative to World" of cell biology is 2.3 and that of immunology is with 1.87 somewhat lower. In other words, both values indicate that documents from this subject categories have a much larger ratio of cites per documents than the world average. We expect lower coefficients for indicators based on raw citation counts if the study were based on subject categories from different disciplines (e.g., materials science or information science). However, there are no peer ratings from disciplines other than the life-sciences available in databases like F1000.

Among the normalized indicators considered here, Percentile in Subject Area (at best) and Category Actual/Expected Citations (also) should be preferred in research assessment studies using bibliometrics. In contrast, Journal Actual/Expected Citations should not be used. However, since (1) the data base of this study is rather small (only a small fraction of papers in cell biology and immunology could be found in F1000) (see Kreiman & Maunsell, 2011), (2) only two subject categories are considered, and (3) the bibliometric indicators are restricted to German papers only, more comprehensive future studies are needed. Until now, there is only one other study available (Li & Thelwall, 2012) with a similar research design. The data set of these authors consisted of 1,397 selected F1000 Genomics & Genetics articles from 2008. The study correlated only the Journal Impact Factor and raw citation counts with FFa, but not field-normalized indicators. Whereas Li and Thelwall (2012) computed with $r=.359$ a very similar coefficient for the correlation between FFa and Journal Impact Factors, the coefficient for the correlation between FFa and WoS citations is different with r=.295 (although this coefficient is also within the confidence interval computed in this study).

# References

Allen, L., Jones, C., Dolby, K., Lynn, D., & Walport, M. (2009). Looking for landmarks: the role of expert review and bibliometric analysis in evaluating scientific publication outputs. *Plos One, 4*(6). doi: 10.1371/journal.pone.0005910.

Bollen, J., Van de Sompel, H., Hagberg, A., & Chute, R. (2009). A Principal Component Analysis of 39 scientific impact measures. *PLoS ONE, 4*(6), e6022.

Bornmann, L. (2010). Towards an ideal method of measuring research performance: some comments to the Opthof and Leydesdorff (2010) paper. *Journal of Inormetrics, 4*(3), 441-443.

Bornmann, L. (2011). Scientific peer review. *Annual Review of Information Science and Technology, 45*, 199-245.

Bornmann, L., & Daniel, H.-D. (2009). Universality of citation distributions. A validation of Radicchi et al.'s relative indicator $c_f = c/c_0$ at the micro level using data from chemistry. *Journal of the American Society for Information Science and Technology, 60*(8), 1664-1670.

Bornmann, L., Mutz, R., & Daniel, H.-D. (2008). Are there better indices for evaluation purposes than the *h* index? A comparison of nine different variants of the *h* index using data from biomedicine. *Journal of the American Society for Information Science and Technology, 59*(5), 830-837. doi: 10.1002/asi.20806.

Bornmann, L., Mutz, R., Marx, W., Schier, H., & Daniel, H.-D. (2011). A multilevel modelling approach to investigating the predictive validity of editorial decisions: do the editors of a high-profile journal select manuscripts that are highly cited after publication? *Journal of the Royal Statistical Society - Series A (Statistics in Society), 174*(4), 857-879. doi: 10.1111/j.1467-985X.2011.00689.x.

Butler, L., & McAllister, I. (2011). Evaluating university research performance using metrics. *European Political Science, 10*(1), 44-58. doi: 10.1057/eps.2010.13.

Cohen, J. (1988). *Statistical power analysis for the behavioral sciences* (2nd ed.). Hillsdale, NJ, USA: Lawrence Erlbaum Associates, Publishers.

Council of Canadian Academies. (2012). *Informing research choices: indicators and judgment: the expert panel on science performance and research funding.* . Ottawa, Canada: Council of Canadian Academies.

F1000. (2012). What is F1000? Retrieved October 25, from http://f1000.com/about/whatis

Figueredo, E. (2006). The numerical equivalence between the impact factor of journals and the quality of the articles. *Journal of the American Society for Information Science and Technology, 57*(11), 1561.

Franceschet, M., & Costantini, A. (2011). The first Italian research assessment exercise: a bibliometric perspective. *Journal of Informetrics, 5*(2), 275-291. doi: DOI: 10.1016/j.joi.2010.12.002.

Garfield, E. (1979). *Citation indexing - its theory and application in science, technology, and humanities*. New York, NY, USA: John Wiley & Sons, Ltd.

Haslam, N., & Laham, S. M. (2010). Quality, quantity, and impact in academic publication. *European Journal of Social Psychology, 40*(2), 216-220. doi: 10.1002/ejsp.727.

InCites[TM] Thomson Reuters. (2012a). *Report created: 03.09.2012. Data processed: 09.05.2012*. Data Source: Web of Science. This data is reproduced under a license from Thomson Reuters. Subject area baseline data processed Jan. 1, 1981->Dec. 31, 2011.

InCites[TM] Thomson Reuters. (2012b). *Report created: 09.07.2012. Data processed: 27.02.2012*. Data Source: Web of Science. This data is reproduced under a license from Thomson Reuters. Subject area baseline data processed Jan. 1, 1981->Dec. 31, 2010.

InCites[TM] Thomson Reuters. (2012c). *Report created: 29.10.2012. Data processed: Dec 31, 2011*. Data Source: Web of Science. This data is reproduced under a license from Thomson Reuters.

Kohler, U., & Kreuter, F. (2012). *Data analysis using Stata* (3. ed.). College Station, TX, USA: Stata Press, Stata Corporation.

Kreiman, G., & Maunsell, J. H. R. (2011). Nine criteria for a measure of scientific output. *Frontiers in Computational Neuroscience, 5*. doi: 10.3389/fncom.2011.00048.

Leydesdorff, L. (2009). How are new citation-based journal indicators adding to the bibliometric toolbox? *Journal of the American Society for Information Science and Technology, 60*(7), 1327-1336.

Li, X., & Thelwall, M. (2012). F1000, Mendeley and traditional bibliometric indicators. In E. Archambault, Y. Gingras & V. Lariviere (Eds.), *The 17th International Conference on Science and Technology Indicators* (pp. 541-551). Montreal, Canada: Repro-UQAM.

Mahdi, S., d'Este, P., & Neely, A. D. (2008). *Citation counts: are they good predictors of RAE scores? A bibliometric analysis of RAE 2001*. London, UK: Advanced Institute of Management Research.

Martin, B. R., & Irvine, J. (1983). Assessing basic research - some partial indicators of scientific progress in radio astronomy. *Research Policy, 12*(2), 61-90.

McKay, S. (2012). Social policy excellence - peer review or metrics? Analyzing the 2008 Research Assessment Exercise in social work and social policy and administration. *Social Policy & Administration, 46*(5), 526-543. doi: 10.1111/j.1467-9515.2011.00824.x.

Medical Research Council. (2009). F1000 evaluation of MRC publication output. Retrieved October 24, from http://www.mrc.ac.uk/Achievementsimpact/Outputsoutcomes/MRCe-Val2009/Publications/index.htm

Rehn, C., Kronman, U., & Wadskog, D. (2007). *Bibliometric indicators – definitions and usage at Karolinska Institutet*. Stickholm, Sweden: Karolinska Institutet University Library.

Riffenburgh, R. H. (2012). *Statistics in medicine* (3. ed.). Oxford, UK: Elsevier.

Ruscio, J., Seaman, F., D'Oriano, C., Stremlo, E., & Mahalchik, K. (in press). Measuring scholarly impact using modern citation-based indices. *Measurement: Interdisciplinary Research and Perspectives*.

Sheskin, D. (2007). *Handbook of parametric and nonparametric statistical procedures* (4th ed.). Boca Raton, FL, USA: Chapman & Hall/CRC.

Smith, A., & Eysenck, M. (2002). *The correlation between RAE ratings and citation counts in psychology*. London: Department of Psychology, Royal Holloway, University of London, UK.

StataCorp. (2011). *Stata statistical software: release 12*. College Station, TX, USA: Stata Corporation.

Vinkler, P. (2010). *The evaluation of research by scientometric indicators*. Oxford, UK: Chandos Publishing.

Wardle, D. A. (2010). Do 'Faculty of 1000' (F1000) ratings of ecological publications serve as reasonable predictors of their future impact? *Ideas in Ecology and Evolution, 3*, 11-15.

Wets, K., Weedon, D., & Velterop, J. (2003). Post-publication filtering and evaluation: Faculty of 1000. *Learned Publishing, 16*(4), 249-258.

Wouters, P., & Costas, R. (2012). *Users, narcissism and control – tracking the impact of scholarly publications in the 21st century*. Utrecht, The Netherlands: SURFfoundation.

Table 1. Description of the variables

| Variable | Number of papers | Mean | Standard deviation | Minimum | Maximum |
| --- | --- | --- | --- | --- | --- |
| Times Cited | 125 | 71.70 | 84.61 | 0 | 489 |
| 2nd Generation Citations | 125 | 592.13 | 946.97 |  | 6176 |
| 2nd Generation Citations per Citing Document | 123 | 6.95 | 5.28 | 0 | 27.72 |
| Journal Actual/Expected Citations | 125 | 2.55 | 3.84 | 0 | 33.33 |
| Category Actual/Expected Citations | 123 | 6.65 | 7.70 | 0 | 45.19 |
| Percentile in Subject Area | 122 | 88.95 | 15.68 | 0 | 99.99 |
| Journal Impact Factor | 125 | 15.38 | 10.27 | 0.91 | 53486 |
| FFa | 125 | 8.34 | 3.29 | 6 | 26 |

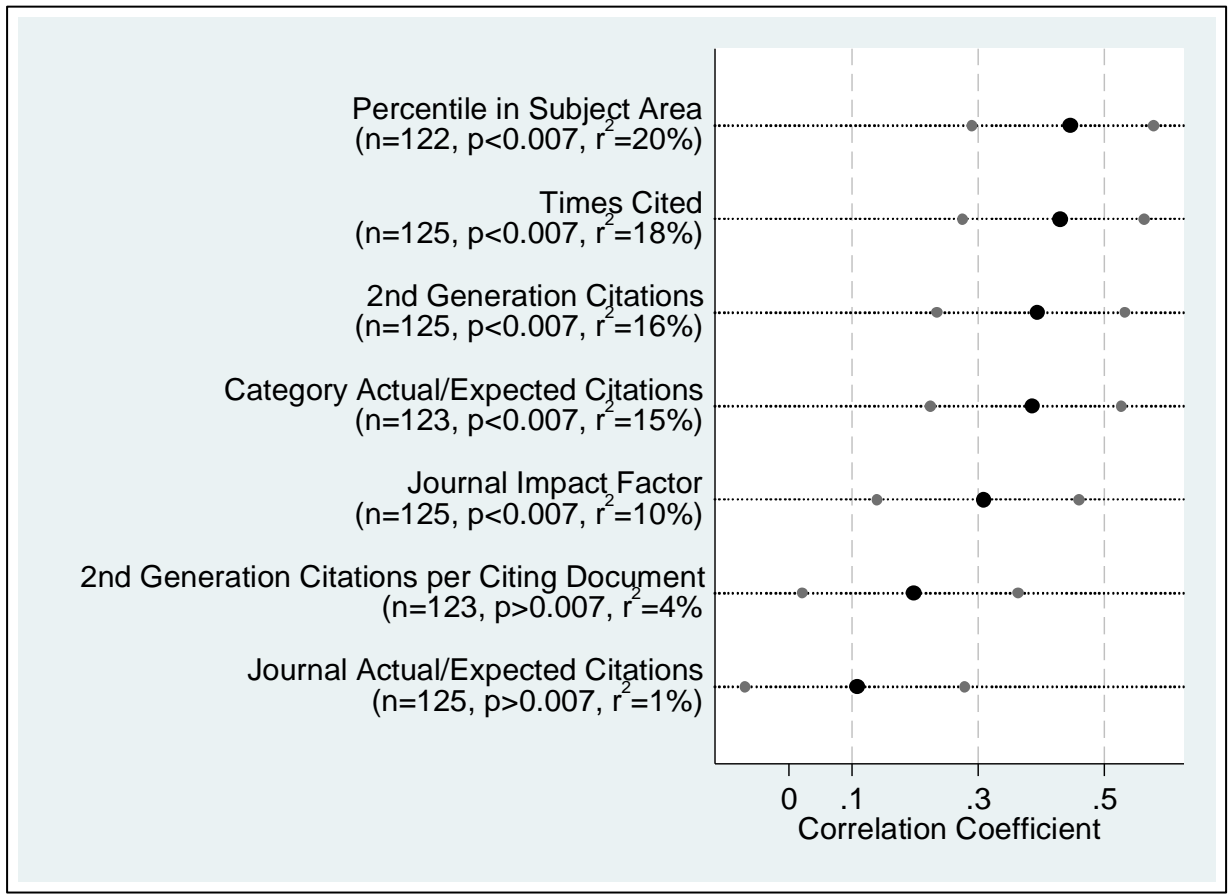

Figure 1. Spearman's rank order correlation coefficients with 95% confidence intervals for the correlation between F1000 Article Factor (FFa) and seven bibliometric metrics. The Bonferroni adjusted α for the statistical test is 0.05/7=0.007 (Riffenburgh, 2012). Coefficients of .1, .3, and .5 are considered by Cohen (1988) as small, middle, and large effects.